\documentclass[9pt,a4paper]{extarticle}
\pdfoutput=1
\usepackage[pdftex]{graphicx}                        
\usepackage[paper=a4paper, margin={1.75cm},centering]{geometry}

\usepackage[utf8]{inputenc} 
\usepackage[T1]{fontenc}    
\usepackage{amsmath}
\usepackage{amsfonts}
\usepackage{mathpazo}
\usepackage{inconsolata}                             
\usepackage{xcolor}
\usepackage[hyphens,obeyspaces,spaces]{url}
\usepackage{hyperref}
\usepackage{multicol}
\usepackage{xspace}                                
\usepackage[detect-all]{siunitx}
\usepackage[font=small]{caption}
\usepackage{bm}
\usepackage{microtype}      
\usepackage{float}			
\usepackage{authblk}

\graphicspath{{figures/}}                           

\newcommand{\subrm}[1]{\ensuremath{_{\text{#1}}}}

\providecommand{\degree}{\ensuremath{^\circ}\xspace}    
\renewcommand{\degree}{\ensuremath{^\circ}\xspace}      
\newcommand{\figref}[1]{Fig.~\ref{#1}}

\newcommand{\vv}[1]{\bm{\mathrm{#1}}}                   

\def\citenum#1{\def\@cite##1##2{##1}\cite{#1}}
\newcommand{\pol}[1]{\textbf{\textsf{#1}}}
\newcommand{\Cdot}{\ensuremath{\boldsymbol{\cdot}}}
\newcommand{\silent}[1]{}

\makeatletter      
   \newcommand{\figcaption}{\captionsetup{type=figure} \def\@captype{figure}\caption}            
   \newcommand{\tabcaption}{\captionsetup{type=table} \def\@captype{table}\caption}            
   \renewcommand{\@biblabel}[1]{#1.}    
\makeatother

\title{Geometric phase evolution}

\author[1\thanks{\tt{nh@hagenlab.org}}]{Nathan Hagen}
\author[1,2]{Luis Garza-Soto}
\affil[1]{Department of Optical Engineering, Utsunomiya University, 7-1-2 Yoto, Utsunomiya, Tochigi 321-8585 Japan}
\affil[2]{Department of Aerospace Structures and Materials, Delft University of Technology, Kluyverweg 1, 2629 HS, Delft, the Netherlands}

\hypersetup{
   pdftitle={Geometric phase evolution},
   pdfauthor={Nathan Hagen, Luis Garza-Soto}
}

\begin{document}

\twocolumn[ 
  \begin{@twocolumnfalse} 
  
\maketitle

\begin{abstract}
   Geometric phase has historically been defined using closed cycles of polarization states, often derived using differential geometry on the Poincar{\'e} sphere. Using the recently-developed wave model of geometric phase, we show that it is better to define geometric phase more generally, allowing every polarized wave to have a well-defined value at any point in its path. Using several example systems, we show how this approach provides more insight into the wave's behavior. Moreover, by tracking the continuous evolution of geometric phase as a wave propagates through an optical system, we encounter a natural explanation of why the conventional Poincar{\'e} sphere solid angle method uses geodesic paths rather than physical paths of the polarization state.
\end{abstract}
\vspace{0.35cm}

  \end{@twocolumnfalse} 
] 

\section{Introduction}

The existing literature defines geometric phase by cycles of states.\cite{Aharonov1987} Since the phase of a wave is gauge-dependent (depends on our choice of basis for the polarization states), the phase value contains an arbitrary offset~\cite{Jisha2021}. However, if we consider cyclical operations that return a polarized wave to its starting state, then the phase remaining after a full cycle is no longer gauge-dependent. The arbitrary offset cancels out regardless of the choice of basis, and what remains has been commonly referred to as the geometric phase~\cite{Giavarini1989}. 

Some researchers have used the existing theoretical framework to define the geometric phase for an open path~\cite{Pati1998,Galvez1999,Dijk2010a,Gutierrez-Vega2011}, basically by determining the proper phase corresponding to ``closing'' the path into a cycle. This too indirectly reinforces the existing framework that it is only possible to talk about geometric phase in terms of cyclical paths. Considering only cycles of states, however, severely limits the information we can obtain about a wave. For example, if geometric phase is only defined for a cycle, and undefined until the cycle is complete, then this limits what we can say about the phase of the wave between start and end. Further, if we consider that in experimental systems no paths are ever exactly closed, then it may be unclear what the closed path requirement even means.

We have previously shown \cite{Garza-Soto2023} that the geometric phase $\gamma$ can be defined such that it has a value for any wave at any point along its propagation through a system --- a quantity that continuously evolves according to its interactions with optical elements and according to coordinate transformations. With this new definition, the change in $\gamma$ between the starting and ending points of any cycle of states, $\Delta \gamma = \gamma\subrm{end} - \gamma\subrm{start}$ is then what the existing literature refers to as the geometric phase~\cite{Berry2024} but which we regard as the \emph{change} in geometric phase. In the discussion below, we show why this generalized definition is useful --- that it provides richer information about the physical behavior of the wave, while retaining the previous cyclical state information as a subset.

The most commonly used method for quantifying the geometric phase for a cyclical sequence of polarization states is to draw the sequence of states on the Poincar{\'e} sphere, connect each state in the cycle using geodesic curves, and calculate the solid angle subtended by the resulting path~\cite{Dijk2010}. The geometric phase ($\gamma$, in radians) is then given by half the solid angle ($\Omega$, in steradians): $\Delta \gamma = -\Omega / 2$, for paths traced in a clockwise sense~\cite{Gutierrez-Vega2011}. For paths traced in an anticlockwise sense, $\Delta \gamma = +\Omega / 2$. This recipe for calculating $\Delta \gamma$, however, also contains a number of restrictions:
\begin{enumerate} \setlength{\itemsep}{-2pt}
   \item The solid angle is calculated by tracing a geodesic curve between the polarization states before entering and after leaving each homogeneous optical element, even when this geodesic is not the actual path followed by the polarization state as it traverses the element. This is the ``geodesic rule''~\cite{Courtial1999,Zhou2020}. 
   \item For the case of inhomogeneous optical elements, such as a cholesteric liquid crystal waveplate (which are linear retarders with an azimuth angle that rotates continuously along the propagation direction), the geodesic rule no longer holds, and the solid angle is calculated using the actual physical path of the polarization state~\cite{Berry1996}.
   \item For cases in which one can chose from among multiple equally-short geodesics in order to close a path, one must choose the one that coincides with the physical path if such a physical path exists~\cite{Palmieri2023}.
\end{enumerate}

The newly developed wave model for geometric phase~\cite{Garza-Soto2023} allows us to explain why these restrictions exist for the solid angle approach. Using the wave model operating on the electromagnetic wave vector obtained from Jones calculus, we validate the often stated principle that geodesics are paths along which $\gamma$ does not accumulate phase~\cite{Palmieri2023,Chryssomalakos2023}. Along the way, we demonstrate that the above rules are incomplete --- that we need to add a new fourth rule:
\begin{enumerate} \setcounter{enumi}{3}
   \item If two adjacent optical elements in the wave's path share the same eigenbasis, then they must be combined into a single element for the purpose of calculating $\Omega$.
\end{enumerate}

In contrast to the differential geometry approach typically used for geometric phase~\cite{Berry1984,Shapere1989,Anandan1988a,Bohm1991,Nityananda2014}, our analysis below uses only the Jones calculus and simple basis transformations so that $\gamma$ retains a clear physical meaning at each step.


\section{Analyzing polarization transformations without propagation phase}

In order to trace the evolving polarization state and geometric phase of a wave propagating through a sequence of optical components, we need to make sure that our model eliminates the effects of propagation phase, leaving only geometric effects. This will allow us to model the smooth change in $\gamma$ as a wave passes through a linear retarder. For the first time, we will be able to see how the geometric phase evolves as it passes through these elements, and not only at the points just before entering and just after leaving an element. As we will see, there can be a lot going on between these two points.

For a wave's polarization state, we use the electric field vector
\begin{equation}\label{eq:Jones_defn}
   \vv{E} = \begin{pmatrix} A_1 e^{i \phi_1} \\ A_2 e^{i \phi_2} \end{pmatrix} \, ,
\end{equation} 
where $A_1$ and $A_2$ are the wave amplitudes, and $\phi_1$ and $\phi_2$ the corresponding wave phases, along each of two orthogonal basis directions 1 and 2. Although the most common choice of basis is $x$-$y$, we will make use of other polarization bases too.

In order to simplify notation when passing through a sequence of standard polarization states, we will use the following letters to represent the corresponding states:
{\setlength{\leftmargini}{15mm}
\begin{itemize} \setlength{\itemsep}{0pt}
   \item[\pol{H}:] horizontally (0\degree linear) polarized
   \item[\pol{V}:] vertically (90\degree linear) polarized
   \item[\pol{D}:] diagonally (45\degree linear) polarized
   \item[\pol{A}:] antidiagonally ($-45\degree$ linear) polarized
   \item[\pol{R}:] right-circularly polarized
   \item[\pol{L}:] left-circularly polarized
   \item[\pol{E}:] any elliptically polarized
\end{itemize}}
On the Poincar{\'e} sphere representation of polarization states, each pair of orthogonal states \pol{HV}, \pol{DA}, and \pol{RL} occur at the opposing points on the sphere along a given axis, as shown in \figref{fig:sequence1}a. To prevent clutter, the figure only shows one of each pair of states along each axis.

In the wave model for geometric phase, the geometric phase $\gamma$ for the addition of any two waves expressed in a linear basis can be written as~\cite{Garza-Soto2023}
\begin{equation}\label{eq:wrapped_gamma}
   \tan (2 \gamma) = \frac{A_1^2 \sin (2 \phi_1) + A_2^2 \sin (2 \phi_2)}{A_1^2 \cos (2 \phi_1) + A_2^2 \cos (2 \phi_2)} \, .
\end{equation}
Any polarized wave can be decomposed into two orthogonal wave components (1 and 2), and so if we consider the addition of wave 1 with its orthogonal component wave 2, then we see that $\gamma$ under this model can be defined for any wave, regardless of its history. In this case, the geometric phase represents the position of the wave peak relative to the axis that we have defined as our phase reference. Orientation angles are defined with respect to the positive $x$-axis, so that a wave has zero phase if it reaches its maximum amplitude when its electric field vector is aligned to the $x$-axis.

When a polarization state passes through a birefringent optical element, the state is transformed according to the element's Jones matrix. For example, the matrix of a linear retarder has the form~\cite{Chipman2019}
\begin{align}
   &\vv{LR} (\theta,\delta,\Phi) = e^{i \Phi} \left[ \vv{R} (-\theta) \begin{pmatrix} e^{-i \delta / 2} &0 \\ 0 &e^{+i \delta / 2} \end{pmatrix} \vv{R} (\theta) \right] \label{eq:Jones_matrix_retarder1} \\
           &= e^{i \Phi} \begin{pmatrix} 
                e^{-i \delta / 2} \cos^2 \theta + e^{i \delta / 2} \sin^2 \theta &(e^{-i \delta / 2} - e^{i \delta / 2}) \cos \theta \sin \theta \\
                (e^{-i \delta / 2} - e^{i \delta / 2}) \cos \theta \sin \theta &e^{i \delta / 2} \cos^2 \theta + e^{-i \delta / 2} \sin^2 \theta
             \end{pmatrix} \label{eq:Jones_matrix_retarder2}
\end{align} 
where $\delta$ is the retardance. The operator $\vv{R} (\theta)$ represents a rotation matrix:
\begin{equation}
   \vv{R} (\theta) = \begin{pmatrix} \cos \theta &\sin \theta \\ -\sin \theta &\cos \theta \end{pmatrix}
\end{equation}
The angle $\theta$ is the orientation angle of the element's axis with respect to the $x$-axis. For a retarder, this is the retarder's fast axis angle. The central matrix in \eqref{eq:Jones_matrix_retarder1} represents a linear retarder in its eigenbasis, so that its fast axis azimuth is oriented along the first eigenstate

In Eqs~\ref{eq:Jones_matrix_retarder1}\,\&\,\ref{eq:Jones_matrix_retarder2}, $\Phi$ is the propagation phase --- the mean phase accumulated by the wave upon propagating through the element. For any given Jones matrix, the propagation phase is calculated as the average of the phases accumulated by each of the eigenstates (the eigenpolarizations). This is equal to the phase accumulated by passing through an equivalent glass plate of the same refractive index:
\begin{equation}
   \Phi = \frac{1}{2} \sum_j 2 \pi n_j \ell / \lambda \, ,
\end{equation} 
for refractive index $n$, plate thickness $\ell$, and wavelength $\lambda$.

Although there are different forms for the Jones matrices, when working with geometric phase it is important to use a definition of the Jones matrices in which the propagation phase can be removed, so that only geometric components remain. Thus, in our work below we set $\Phi = 0$, and allow the phases to operate symmetrically on the element's eigenstates --- the ``symmetric phase convention''~\cite{Chipman2019}.

An important property of the rotation and retarder Jones matrices that we will use is that their parameters are additive on matrix concatenation. That is, if we concatenate $N$ retarders, each with $\delta / N$ retardance, then the concatenated system will be equivalent to a single retarder matrix with retardance $\delta$:
\begin{equation}\label{eq:retarder_slicing}
   \vv{LR} (\theta, \delta) = \prod_n \vv{LR} (\theta, \delta / N) \, ,
\end{equation}
and the same additive property holds for rotation matrices:
\begin{equation}\label{eq:rotator_slicing}
   \vv{R} (\theta) = \prod_n \vv{R} (\theta / N) \, .
\end{equation}
Thus, if we want to know how the polarization state evolves as it propagates inside a linear retarder, then we can split the retarder into many thin retarder elements, and analyze the polarization state after passing through each thin element. In the limit of $N \to \infty$, the matrices represent differential elements~\cite{Jones1948,Berry1996}.

\section{Tracking geometric phase along a path using the Jones calculus}\label{sec:sequence1}

\begin{figure*}
   \centering
   \includegraphics[width=0.78\linewidth]{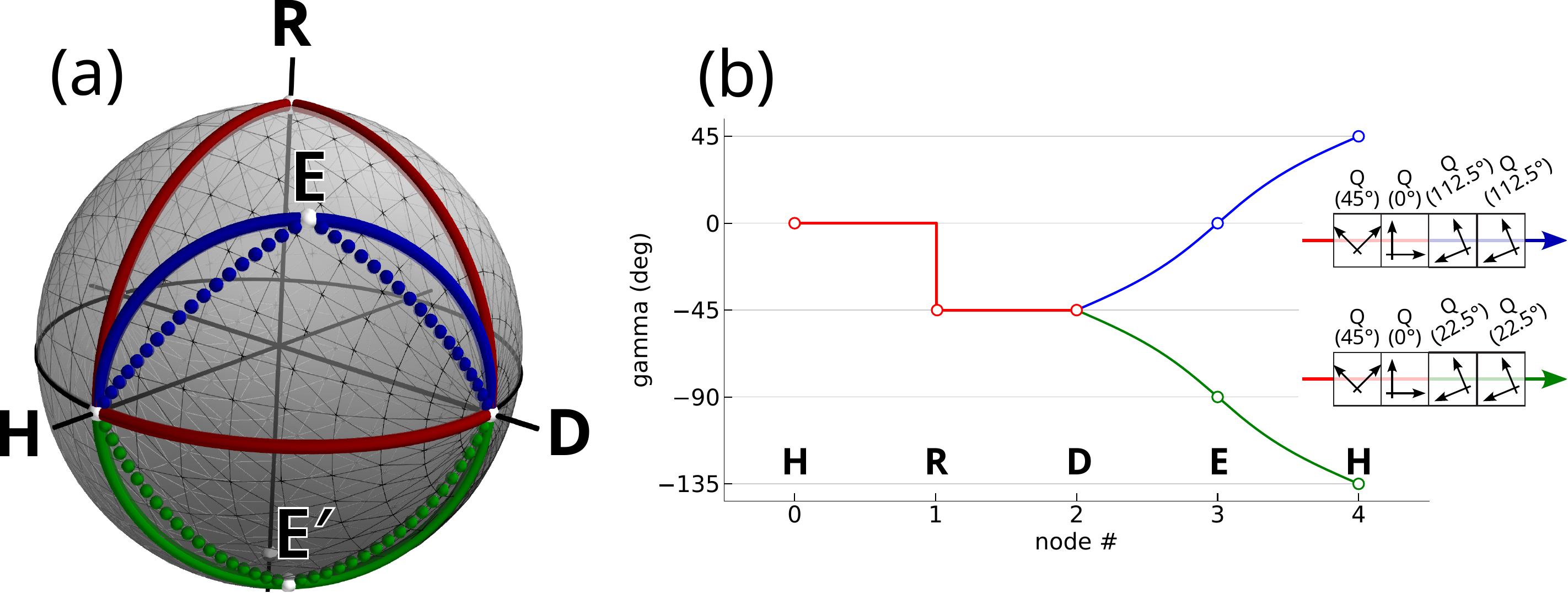}
   \caption{(a) Sequence \#1: \pol{H}-\pol{R}-\pol{D}-\pol{H}. The first two (red) path segments are performed with QWPs oriented at 45\degree and 0\degree, respectively. The third segment of the cyclical path is performed using (blue) two QWPs with fast axis at 112.5\degree, or (green) two QWPs with fast axis at 22.5\degree. The dotted green and blue paths show the of geodesic curves from \pol{D} to \pol{E} (the midpoint) and then from \pol{E} to \pol{H}. (b) The calculated geometric phase for both paths, using \eqref{eq:wrapped_gamma} in the \pol{HV} basis. The green path is unwrapped at the $-\pi/2$ boundary in order to clearly illustrate its symmetry with the blue path. (b, inset) The two different optical systems used for the two paths modeled.}
   \label{fig:sequence1}
\end{figure*}

To illustrate our method for tracking geometric phase evolution, we first consider transformations of a wave with horizontal polarization state \pol{H} passing through an optical system, and we analyze the geometric phase $\gamma$ as it propagates through this system. The system consists of four quarter-wave plate (QWP) elements: one QWP with fast axis oriented at 45\degree, followed by a second QWP oriented at 0\degree, and finally by a pair of QWPs that are either oriented at 112.5\degree (blue path) or 22.5\degree (green path). The resulting path of polarization states are drawn on the Poincar{\'e} sphere in \figref{fig:sequence1}(a). The geometric phase at each point along the polarization state's evolution as it passes through this system is shown in \figref{fig:sequence1}(b). The red path shows the polarization state starting at \pol{H}, rotating up to the pole at \pol{R}, rotating again to \pol{D}, and finally returning to \pol{H} along either the blue or green curves. Each nonadiabatic change in a path is considered a node, represented by a small white dot. However, as we will see below, the node placed halfway along the blue and green curves, indicating the state between the two QWPs, is an adiabatic point. If we temporarily ignore this node, as we would do if we used HWPs in place of pairs of QWPs, then we could observe that even though the blue and green paths traverse different physical polarization paths, they traverse the same sequence of nodes.

In the common approach used by the existing literature, the geometric phase produced by traversing this path is calculated by connecting the various nodes by geodesic arcs. The solid angle $\Omega$ subtended by the area inside the curve, leads to $\gamma = -\Omega / 2$~\cite{Garza-Soto2023a}. In one of the example paths shown in \figref{fig:sequence1}, the node-connected path segments trace a spherical triangle shown by three red curves. This triangle encloses a solid angle of $\Omega = \pi / 2$ steradians, giving $\gamma = -\pi/4$. (In order to clearly differentiate the various phases and angles, orientation angles will always be stated in degrees, while geometric phases will be given in radians. An exception is made for the figures, where using degrees for $\gamma$ is more natural.)

Next we consider the same path, calculated instead using the wave model. For this approach, we create an input Jones vector representing the horizontal polarization state, in the canonical $x$-$y$ basis: $\pol{H} = \big( \begin{smallmatrix} 1 \\ 0 \end{smallmatrix} \big)$. We then split each of the optical components into a series of thin elements and record the polarization state at each step as it passes through the system. Thus, we obtain a sequence of polarization states representing steps along the curves between nodes and also the nodes themselves. For each state, we calculate $\gamma$ using \eqref{eq:wrapped_gamma}, with the result shown in \figref{fig:sequence1}(b). From the figure, we can see that the phase starts at zero at state \pol{H}, suddenly decreases to $-45\degree$ as it leaves \pol{R}, and stays at $-45\degree$ as it reaches \pol{D}, but after this splits in two directions depending on whether the wave travels along the blue or green path, reaching either $+45\degree$ or $-135\degree$ at the end.

In order to see how the geometric phase arises from the various operations, we can analyze the wave phase directly from the polarization state vector. At each optical element, we make sure that the polarization state transformation matrix contains no propagation phase, so that any changes to the state are purely due to geometric effects.

The path going from the initial state \pol{H} towards \pol{R} uses a QWP with fast axis oriented at 45\degree, which means that the eigenbasis for this QWP is the pair of states \pol{D}\,\&\,\pol{A}. If we represent the horizontal input state in the \pol{DA} basis, then we have
\begin{equation}\label{eq:R}
   \vv{E}_{\pol{DA}} = \frac{1}{\sqrt{2}} \begin{pmatrix} 1 &1 \\ -1 &1 \end{pmatrix} \Cdot \vv{E}_{\pol{HV}} = \frac{1}{\sqrt{2}} \begin{pmatrix} 1 \\ -1 \end{pmatrix} \, .
\end{equation}
where $\vv{E}_{\pol{HV}}$ is the input electric field vector represented in the \pol{HV} basis, and $\vv{E}_{\pol{DA}}$ the same field represented in the \pol{DA} basis.

In the eigenbasis, the Jones matrix of the retarder is simply
\begin{equation}\label{eq:LR_diagonal}
   \vv{LR} (\delta, 0) = \begin{pmatrix} e^{-i \delta / 2} &0 \\ 0 &e^{+ i \delta / 2} \end{pmatrix} \, ,
\end{equation}
which clearly contains no propagation phase. Applying this matrix to the input state produces
\begin{equation}
   \vv{E}'_{\pol{DA}} = \begin{pmatrix} e^{-i \delta / 2} &0 \\ 0 &e^{+ i \delta / 2} \end{pmatrix} \Cdot \frac{1}{\sqrt{2}} \begin{pmatrix} 1 \\ -1 \end{pmatrix} = \frac{1}{\sqrt{2}} \begin{pmatrix} e^{-i \delta / 2} \\ -e^{+i \delta / 2} \end{pmatrix} \, ,
\end{equation}
for retardance $\delta$. The state of polarization slowly evolves as the wave passes through the retarder, eventually reaching $\delta = \pi / 2$ for a QWP, so that polarization state exiting the retarder becomes
\begin{equation}
   \vv{E}'_{\pol{DA}} = \frac{1}{\sqrt{2}} \begin{pmatrix} e^{-i \pi / 4} \\ e^{-i 3 \pi / 4} \end{pmatrix} = \frac{e^{-i \pi / 4}}{\sqrt{2}} \begin{pmatrix} 1 \\ -i \end{pmatrix} \, .
\end{equation} 
This is the state \pol{R} represented in the \pol{DA} basis.

In the next step of the sequence, going from \pol{R} to \pol{D}, the light passes into a QWP with fast axis oriented at 0\degree. In this case, the eigenbasis is \pol{HV}, and so we translate our Jones vector back into the canonical basis:
\begin{equation}\label{eq:rcp}
   \vv{E}_{\pol{HV}} = \frac{1}{\sqrt{2}} \begin{pmatrix} 1 &-1 \\ 1 &1 \end{pmatrix} \Cdot \vv{E}'_{\pol{DA}} = \frac{1}{\sqrt{2}} \begin{pmatrix} 1 \\ -i \end{pmatrix} \, .
\end{equation}
Since the two components of the polarization vector have the same amplitude, the geometric phase will be halfway between the two vector component phases, so that now $\gamma = -\pi/4$, in agreement with the curve drawn in \figref{fig:sequence1}(c). In this basis, the second retarder's matrix is given by \eqref{eq:LR_diagonal}, so that the curve between nodes \pol{R} and \pol{D}, expressed in the \pol{HV} basis, is
\begin{equation}
   \vv{E}'_{\pol{HV}} = \begin{pmatrix} e^{-i \delta / 2} &0 \\ 0 &e^{+ i \delta / 2} \end{pmatrix} \Cdot \frac{1}{\sqrt{2}} \begin{pmatrix} 1 \\ -i \end{pmatrix} = \frac{1}{\sqrt{2}} \begin{pmatrix} e^{-i \delta / 2} \\ -i e^{+i \delta / 2} \end{pmatrix} \, .
\end{equation} 
Substituting $\delta = \pi / 2$ for the QWP gives the polarization state at the exit surface of the retarder as
\begin{equation}\label{eq:nodeD}
   \vv{E}'_{\pol{HV}} = \frac{e^{-i \pi / 4}}{\sqrt{2}} \begin{pmatrix} 1 \\ 1 \end{pmatrix} \, .
\end{equation} 
Although the electric field vector itself has zero phase, the global phase factor shifts the entire wave by $-\pi/4$. Thus, in the \pol{HV} basis, we can see that the geometric wavefront for state \pol{D} is $\gamma = -\pi/4$. Once again, by viewing the transformation from the perspective of the eigenbasis (the geodesic curve's axis of symmetry), we find that the phase is unchanged. (Note, however, that since our polarization state has changed from our original \pol{H}, the phase measured by an interferometer in this case will not be equal to $\gamma$~\cite{Garza-Soto2023}.)

Now that we have reached state \pol{D}, the final segment of the path involves using a pair of QWPs to rotate the state from \pol{D} back to its original state \pol{H}. Starting with the polarization state at node \pol{D}, we translate the state into the eigenbasis of the retarder (in this case, linear polarization oriented at either 112.5\degree or 22.5\degree) and apply the retardance \eqref{eq:Jones_matrix_retarder1}. Starting with the blue path drawn in \figref{fig:sequence1}, we choose the QWPs oriented at 112.5\degree. We translate the polarization vector at \pol{D} into this retarder's eigenbasis as:
\begin{align}
   \vv{E}_{112} &= \begin{pmatrix} \cos \big( \frac{5\pi}{8} \big) &\sin \big( \frac{5\pi}{8} \big) \\ -\sin \big( \frac{5\pi}{8} \big) &\cos \big( \frac{5\pi}{8} \big) \end{pmatrix} \Cdot \frac{e^{-i\pi/4}}{\sqrt{2}} \begin{pmatrix} 1 \\ 1 \end{pmatrix} \notag \\
      &= \frac{e^{-i\pi/4}}{\sqrt{2}} \begin{pmatrix} \cos \big( \frac{5\pi}{8} \big) + \sin \big( \frac{5\pi}{8} \big) \\ -\sin \big( \frac{5\pi}{8} \big) + \cos \big( \frac{5\pi}{8} \big) \end{pmatrix} \, .
\end{align}
Next we apply the retardance. For a HWP in its eigenbasis, the retarder matrix has the simple form of $\vv{LR} = \big( \begin{smallmatrix} -i &0 \\ 0 &i \end{smallmatrix} \big)$ so that the polarization vector becomes
\begin{align}
   \vv{E}'_{112} &= \begin{pmatrix} -i &0 \\ 0 &i \end{pmatrix} \Cdot \frac{e^{-i\pi/4}}{\sqrt{2}} \begin{pmatrix} \cos \big( \frac{5\pi}{8} \big) + \sin \big( \frac{5\pi}{8} \big) \\ -\sin \big( \frac{5\pi}{8} \big) + \cos \big( \frac{5\pi}{8} \big) \end{pmatrix} \notag \\
      &= \frac{i e^{-i\pi/4}}{\sqrt{2}} \begin{pmatrix} -\cos \big( \frac{5\pi}{8} \big) - \sin \big( \frac{5\pi}{8} \big) \\ \cos \big( \frac{5\pi}{8} \big) - \sin \big( \frac{5\pi}{8} \big) \end{pmatrix} \, . \label{eq:112prime}
\end{align} 
Finally, translating this back into the \pol{HV} basis gives
\begin{equation}\label{eq:path1blue}
   \vv{E}_{\pol{HV}} = \begin{pmatrix} \cos \big( \frac{-5\pi}{8} \big) &\sin \big( \frac{-5\pi}{8} \big) \\ -\sin \big( \frac{-5\pi}{8} \big) &\cos \big( \frac{-5\pi}{8} \big) \end{pmatrix} \Cdot \vv{E}'_{112} = e^{+i\pi/4} \begin{pmatrix} 1 \\ 0 \end{pmatrix} \, .
\end{equation}
This is the result for the blue path. Following a similar procedure for the green path, using $\theta = 22.5\degree$ in place of the above $\theta = 112.5\degree$, we get the result
\begin{equation}\label{eq:path1green}
   \vv{E}_{\pol{HV}} = e^{-i 3 \pi / 4} \begin{pmatrix} 1 \\ 0 \end{pmatrix} \, .
\end{equation} 
In both cases, the polarization state vectors have no phase, so that the global phase factor by itself determines the phase shift. We can see that there is a $\Delta \gamma = \pi$ difference in geometric phase between the blue and green paths, matching the $\gamma$ curves drawn in \figref{fig:sequence1}(b). This phase shift between the blue and green paths typically goes unnoticed.

According to the standard theory, the physical path traversed between nodes has no effect on the geometric phase calculation, but in the situation of the blue and green curves of \figref{fig:sequence1}, we see that it can be ambiguous about which geodesic to choose. The evolution curves clearly show that the two should be $\pi$ radians out of phase, which indicates that the solid angle approach in this case should choose the longer geodesic between the nodes --- the one following the equator along the back side of the sphere --- and not the shorter geodesic that most would require. Evidently a fifth rule is needed to remove this ambiguity, though we have not yet formulated it. However, the fact that $\gamma$ emerges $\pi$ radians out of phase between the blue and green paths is physically meaningful. The two paths experience the final HWP with the fast and slow axes swapped, and there is a $\pi$ retardance difference between the two waveplate orientations. Moreover, the evolution curves for $\gamma$ allow one to visualize the physical behavior as the wave propagates through the crystal: the two eigenwaves are being pushed farther apart in phase from one another as they propagate further through their corresponding retarder elements, causing a shift in the wave peak that is nonlinearly related to their phase separation.

Another observation we can make about the paths drawn in \figref{fig:sequence1}(a) is that since there are two QWPs here rather than a single HWP, the current geodesic rule would require one to draw a geodesic connecting the states at the entrance and exit of the first QWP, and another geodesic at the entrance and exit of the second QWP. These are the paths drawn with dotted green and blue geodesic curves in \figref{fig:sequence1}(a). However, if we incorporate these intermediate geodesic curves in our calculation of the solid angle, we obtain $\Omega\subrm{blue} = \SI{0.892}{rad}$ and $\Omega\subrm{green} = \SI{2.25}{rad}$, rather than the known correct values of $\Omega = \pi/2$. Thus, in order to get the correct answer using the geodesic rule, we instead need to add a new rule (\#4): if the path includes multiple adjacent elements that share an eigenbasis, then these elements must be combined into a single entity, so that the geodesic curve is therefore drawn from the entering state to exit state of that entire subsystem, rather than between each individual optical element. 
In the wave description approach, these special rules are needed.


\section{Why the geodesic rule works}\label{sec:five_paths}

\begin{figure*}
   \centering
   \includegraphics[width=0.65\linewidth]{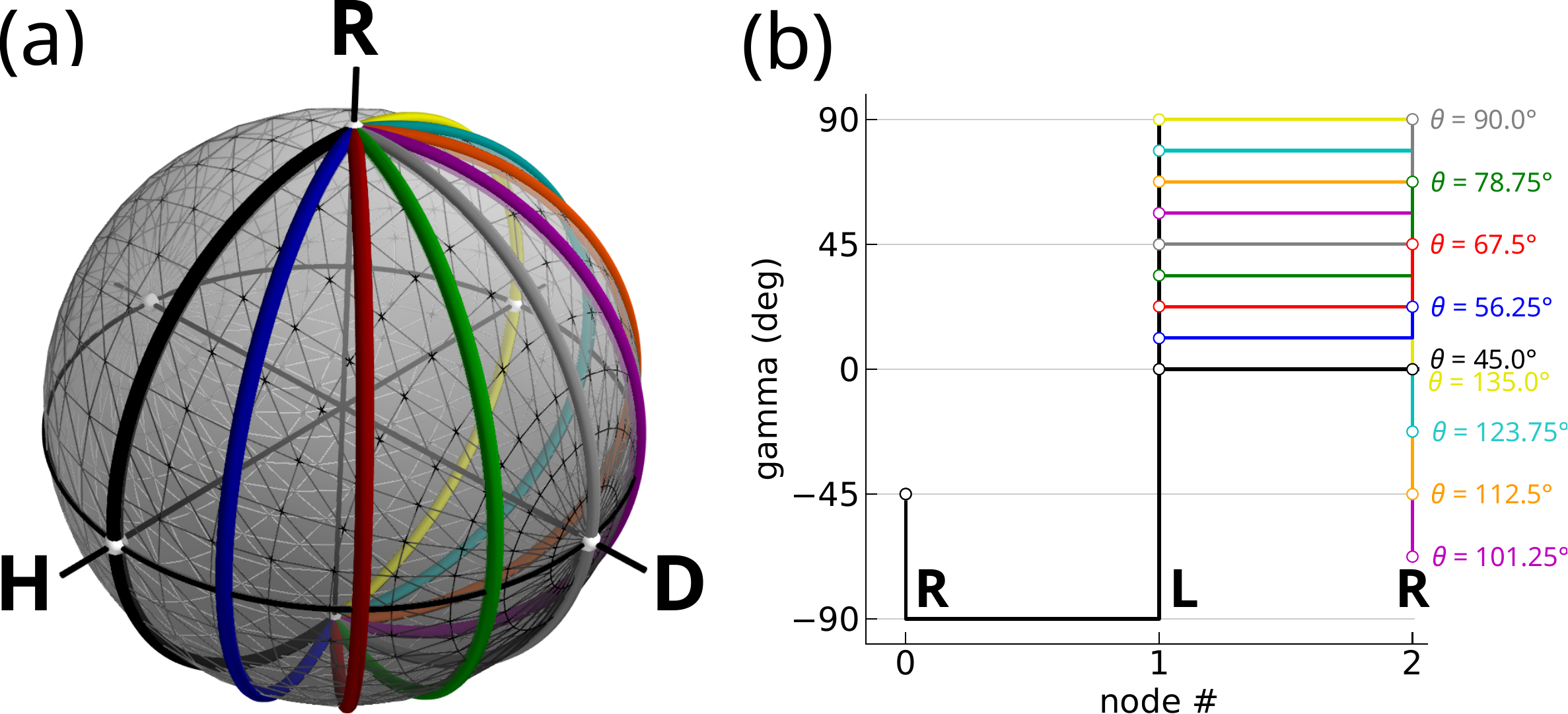}
   \caption{Sequence \#2: (a) Path \pol{R}-\pol{L}-\pol{R}. The first segment from \pol{R} to \pol{L} is performed with a HWP oriented at $-45\degree$. The second segment returning from \pol{L} to \pol{R} is performed with any one of HWPs whose fast axes are oriented at 11.25\degree intervals starting from $+45\degree$. (b) The calculated geometric phase for each path.}
   \label{fig:sequence2}
\end{figure*}

Figure~\ref{fig:sequence2}(a) shows another sequence of polarization states traced on the Poincar{\'e} sphere. As in Sec.~\ref{sec:five_paths}, the paths are generated by slicing each polarization component into hundreds of thin elements, and calculating the polarization state exiting each thin element. In \figref{fig:sequence2}(a), we start with a right-circular polarization state \pol{R} and use a HWP oriented at $\theta = -45\degree$ to convert this into \pol{L}. From there we use a second HWP to return to \pol{R}, but choose a variety of different orientation angles to do so. The figure shows paths for HWPs with fast-axis orientation angles $\theta$ from $+45\degree$ to $+135\degree$ at intervals of 11.25\degree. Figure~\ref{fig:sequence2}(b) shows the corresponding evolution curves for $\gamma$ calculated from \eqref{eq:wrapped_gamma} --- the geometric phase calculated from the perspective of the \pol{HV} basis. 

Following the polarization state evolution with the Jones calculus, we start with the state at \pol{R} written in the \pol{HV} basis,
\begin{equation}
   \vv{E}_{\pol{HV}} = \frac{1}{\sqrt{2}} \begin{pmatrix} 1 \\ -i \end{pmatrix} \, ,
\end{equation} 
which has $\gamma = -\pi/4$. The first HWP has its fast axis oriented at $-45\degree$, so that we start by projecting the above state onto the antidiagonal-diagonal \pol{AD} basis:
\begin{equation}
   \vv{E}_{\pol{AD}} = \begin{pmatrix} \cos \big( \frac{-\pi}{4} \big) &\sin \big( \frac{-\pi}{4} \big) \\ -\sin \big( \frac{-\pi}{4} \big) &\cos \big( \frac{-\pi}{4} \big) \end{pmatrix} \Cdot \frac{1}{\sqrt{2}} \begin{pmatrix} 1 \\ -i \end{pmatrix} = \frac{1}{2} \begin{pmatrix} 1+i \\ 1-i \end{pmatrix} \, .
\end{equation}
(While all of these operations can be performed in a constant basis, switching to each element's eigenbasis provides a clearer analysis.) Note that the \pol{AD} basis differs from the \pol{DA} basis in that the former is rotated by 90\degree with respect to the latter, as needed in order to keep the first polarization basis vector oriented along the waveplate's fast axis.

Now that we are in the HWP's eigenbasis, applying its retardance to the state gives
\begin{equation}
   \vv{E}'_{\pol{AD}} = \begin{pmatrix} -i &0 \\ 0 &i \end{pmatrix} \Cdot \frac{1}{2} \begin{pmatrix} 1+i \\ 1-i \end{pmatrix} = \frac{1}{2} \begin{pmatrix} 1-i \\ 1+i \end{pmatrix} \, .
\end{equation} 
The two components of this state, $1-i$ and $1+i$, correspond to phases of $+45\degree$ and $-45\degree$. Since the amplitudes of the two are equal and the two phases are symmetric about zero, \eqref{eq:wrapped_gamma} gives a geometric phase of zero for this state in this basis.

In the next step, we use a HWP whose orientation angle $\theta$ is not fixed, but we can use the angle as a variable in our basis transformation of the polarization state to obtain
\begin{equation}
   \vv{E}_{\theta} = \begin{pmatrix} \cos \big( \theta + \frac{\pi}{4} \big) &\sin \big( \theta + \frac{\pi}{4} \big) \\ -\sin \big( \theta + \frac{\pi}{4} \big) &\cos \big( \theta + \frac{\pi}{4} \big) \end{pmatrix} \Cdot \vv{E}'_{\pol{AD}} = \frac{e^{i \theta}}{\sqrt{2}} \begin{pmatrix} 1 \\ i \end{pmatrix} \, .
\end{equation}
Here we see that the state acquires a phase of $\theta + \tfrac{\pi}{4}$ as a result of the basis transformation. Now that we are in the HWP's eigenbasis, we apply its retardance with
\begin{equation}
   \vv{E}'_{\theta} = \begin{pmatrix} -i &0 \\ 0 &i \end{pmatrix} \Cdot \frac{e^{i \theta}}{\sqrt{2}} \begin{pmatrix} 1 \\ i \end{pmatrix} 
   = \frac{e^{i \theta}}{\sqrt{2}} \begin{pmatrix} -i \\ -1 \end{pmatrix} \, .
\end{equation} 
Finally, we project this last state onto the \pol{HV} basis:
\begin{equation}
   \vv{E}_{\pol{HV}} = \begin{pmatrix} \cos (-\theta) &\sin (-\theta) \\ -\sin (-\theta) &\cos (-\theta) \end{pmatrix} \Cdot \vv{E}'_{\theta} 
      = \frac{-i e^{i 2 \theta}}{\sqrt{2}}  \begin{pmatrix} 1 \\ -i \end{pmatrix} \, .
\end{equation}
We see that if the return path uses a HWP oriented at $+45\degree$, then the wave phase will be the same as our starting phase. This is as we expect, since this paths subtends no solid angle. However, if the return path uses a HWP oriented at 90\degree, then we find that the wave phase has shifted by $+90\degree$ from our starting point, in agreement with the solid angle approach to calculating $\gamma$, namely that $\Delta \gamma = 2 \theta$.

This procedure reinforces the often-stated feature of geometric phase that $\gamma$ does not accumulate along geodesic paths. Indeed, in the case of \figref{fig:sequence2}, where the two components of the polarization state always keep the same amplitude in the bases that we use, all of the geometric phase is produced by the basis transformations themselves. This points to one way of interpreting the ``geodesic rule'' in the solid angle approach to calculating $\gamma$. The reason we need to link nodes in the path using geodesic curves rather than physical curves, in the case of \figref{fig:sequence2} at least, is that all of the action occurs at the nodes. Viewed in the eigenbasis of a retarder, the polarization state may evolve as it propagates through the retarder, but the geometric phase does not. The separation of the two eigenwaves as they emerge from the retarder, as transformed by any change of basis, induces the shift in wave phase with respect to the phase reference plane.

The special thing about homogeneous optical elements (whose physical paths are to be replaced by geodesics under the geodesic rule) is that the eigenbasis does not change between the point where we enter and leave the element or set of elements. For the geodesic rule to work, we must delay evaluating the polarization state until we emerge from the eigenbasis and are ready to transform the state to a new basis --- as indicated by a node in the path.

\section{An example of different physical paths producing the same $\gamma$}\label{sec:sequence4}

\begin{figure*}
   \centering
   \includegraphics[width=0.75\linewidth]{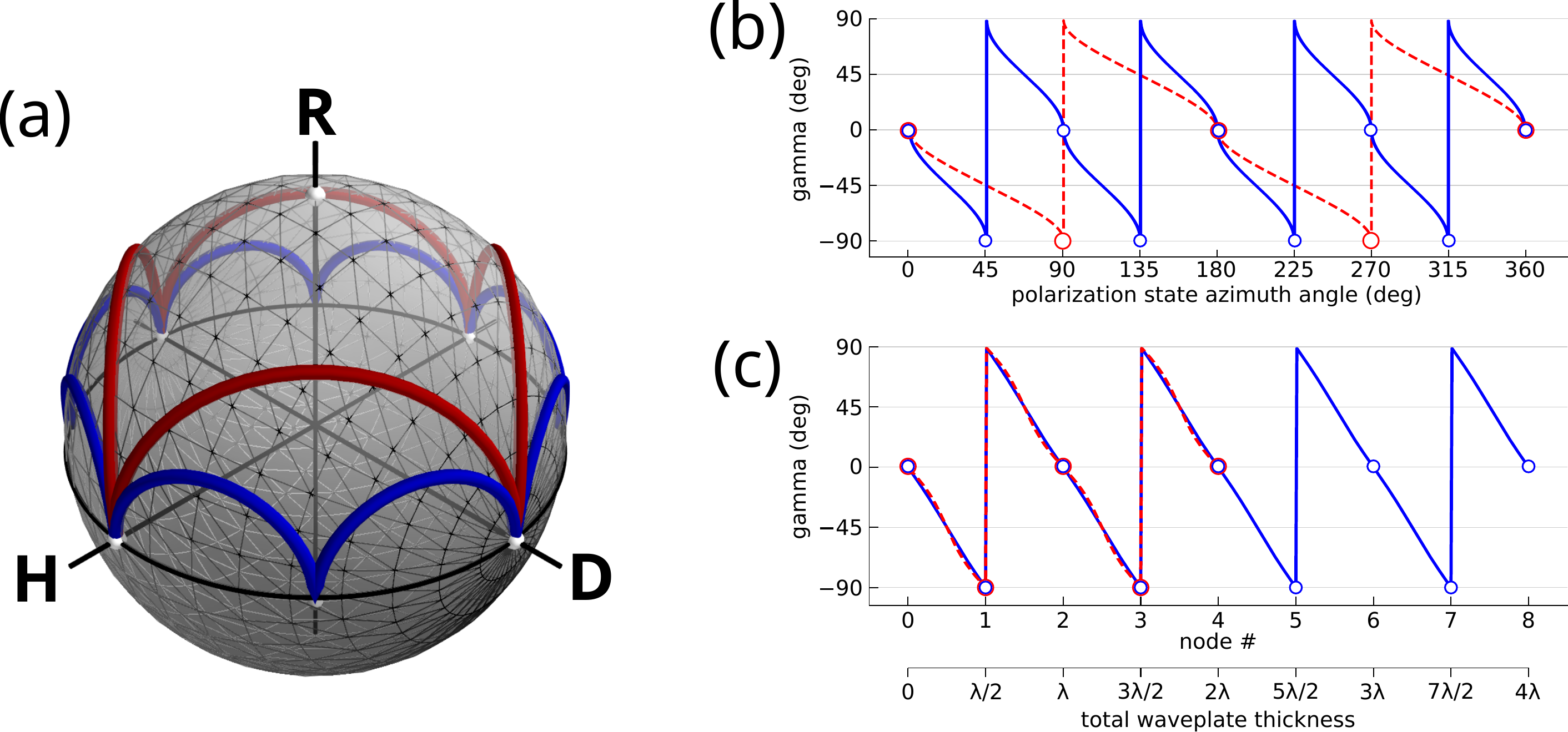}
   \caption{Sequence \#3: Two paths that use HWPs to cycle around the Poincar{\'e} sphere --- one (shown in red) where the orientation angle of each HWP increments by 45\degree at each node, and another (shown in blue) where the HWP angles increment by 22.5\degree. The starting state for each path is \pol{H}. (b,c) The calculated geometric phase for the paths, where (b) and (c) display the same data using different abscissae.}
   \label{fig:sequence3}
\end{figure*}

Figure~\ref{fig:sequence3} shows a pair of paths created with a series of HWPs to loop the polarization state around the Poincar{\'e} sphere. In one path, each successive HWP has its fast axis rotated by 45\degree with respect to the previous HWP. In the second path, the fast axis is incremented by 22.5\degree, creating smaller loops on the sphere. In the conventional solid angle approach to analyzing these paths, the curves connecting the path nodes will all lie along the equator, rather than following the physical polarization state. And in the conventional approach, it is only once the paths return to their starting point, when they close the cycle, that the geometric phase can finally be defined. For both of these paths, this $\Delta \gamma$ will be 0 or $\pi$ because all of the nodes on the path lie on the same great circle. However, in the evolution curves of \figref{fig:sequence1}(b), we can see that there is a lot of activity occurring in the geometric phase along these paths, but that once we return to state \pol{H} we get $\Delta \gamma = \gamma\subrm{end} - \gamma\subrm{start} = n \pi$ for $n \in \{0, 2, 4\}$, in agreement with conventional theory. (In this case, if we are using phase wrapping then $n=0$ for both paths. If unwrapping the phase, then we find that $n=2$ for the red path, and $n=4$ for the blue path.)

The $\gamma$ evolution curves of \figref{fig:sequence3} show that the blue curves exhibit phase changes that are twice as fast as those of the red curves. Thus, for example, the first node of the red curve lies at \pol{D}, which coincides with the polarization state at the second blue curve node. If we draw a geodesic between the stating and ending points, we would use the exact same geodesic for the two curves, but the $\gamma$ of the blue curve will be twice the red curve value there. This should probably not surprise, since the blue curves and red ones both use HWPs, and there are twice as many blue ones, so that the blue path will experience twice as much retardance splitting of waves. One can also note that there are small differences in the shapes of the red and blue curves between nodes, with the red curves showing a larger deviation from linearity.

Following the polarization state evolution with the Jones calculus, we start with the state at \pol{H} and, following the red curve, project the state onto the first HWP's eigenbasis (fast axis oriented at 22.5\degree):
\begin{equation}
   \vv{E}_{22} = \begin{pmatrix} \cos \big( \frac{\pi}{8} \big) &\sin \big( \frac{\pi}{8} \big) \\ -\sin \big( \frac{\pi}{8} \big) &\cos \big( \frac{\pi}{8} \big) \end{pmatrix} \begin{pmatrix} 1 \\ 0 \end{pmatrix} = \begin{pmatrix} \cos \big( \frac{\pi}{8} \big)  \\ -\sin \big( \frac{\pi}{8} \big) \end{pmatrix} \, .
\end{equation}
Applying the halfwave retardance in this basis gives
\begin{equation}
   \vv{E}'_{22} = \begin{pmatrix} -i &0 \\ 0 &i \big) \end{pmatrix} \begin{pmatrix} \cos \big( \frac{\pi}{8} \big)  \\ -\sin \big( \frac{\pi}{8} \big) \end{pmatrix} = -i \begin{pmatrix} \cos \big( \frac{\pi}{8} \big)  \\ \sin \big( \frac{\pi}{8} \big) \end{pmatrix} \, .
\end{equation}
Evaluating this state in the \pol{HV} basis, we obtain
\begin{equation}
   \vv{E}'_{\pol{HV}} = \begin{pmatrix} \cos \big( \frac{-\pi}{8} \big) &\sin \big( \frac{-\pi}{8} \big) \\ -\sin \big( \frac{-\pi}{8} \big) &\cos \big( \frac{-\pi}{8} \big) \end{pmatrix} \Cdot -i \begin{pmatrix} \cos \big( \frac{\pi}{8} \big)  \\ \sin \big( \frac{\pi}{8} \big) \end{pmatrix} = -i \begin{pmatrix} 1 \\ 1 \end{pmatrix} \, ,
\end{equation}
which has a phase shift of $\gamma = -\pi/2$, in agreement with \figref{fig:sequence3}(b), where the 180\degree has been wrapped to 0\degree. Since these curves are smooth, we can easily unwrap them to obtain the unwrapped phase if we wish.

Following the same process for the blue curve, the first HWP's eigenbasis is determined by its fast axis oriented at 11.25\degree. Projecting the state \pol{H} onto this basis gives
\begin{equation}
   \vv{E}_{11} = \begin{pmatrix} \cos \big( \frac{\pi}{16} \big) &\sin \big( \frac{\pi}{16} \big) \\ -\sin \big( \frac{\pi}{16} \big) &\cos \big( \frac{\pi}{16} \big) \end{pmatrix} \begin{pmatrix} 1 \\ 0 \end{pmatrix} = \begin{pmatrix} \cos \big( \frac{\pi}{16} \big)  \\ -\sin \big( \frac{\pi}{16} \big) \end{pmatrix} \, .
\end{equation}
Applying the halfwave of retardance in this basis,
\begin{equation}
   \vv{E}'_{11} = \begin{pmatrix} -i &0 \\ 0 &i \end{pmatrix} \begin{pmatrix} \cos \big( \frac{\pi}{16} \big) \\ -\sin \big( \frac{\pi}{16} \big) \end{pmatrix} = -i \begin{pmatrix} \cos \big( \frac{\pi}{16} \big)  \\ \sin \big( \frac{\pi}{16} \big) \end{pmatrix} \, .
\end{equation}
Next, we rotate this by 22.5\degree to project this state onto the 33.75\degree:123.75\degree basis:
\begin{align}
   \vv{E}_{33} &= \begin{pmatrix} \cos \big( \frac{\pi}{8} \big) &\sin \big( \frac{\pi}{8} \big) \\ -\sin \big( \frac{\pi}{8} \big) &\cos \big( \frac{\pi}{8} \big) \end{pmatrix} \Cdot -i \begin{pmatrix} \cos \big( \frac{\pi}{16} \big)  \\ \sin \big( \frac{\pi}{16} \big) \end{pmatrix} \notag \\
   &= -i \begin{pmatrix} \cos \frac{\pi}{8} \cos \frac{\pi}{16} + \sin \frac{\pi}{8} \sin \frac{\pi}{16} \\ -\cos \frac{\pi}{8} \sin \frac{\pi}{16} + \sin \frac{\pi}{8} \cos \frac{\pi}{16} \end{pmatrix} \, .
\end{align}
Next we apply the halfwave retardance in this basis:
\begin{equation}
   \vv{E}'_{33} = \begin{pmatrix} -i &0 \\ 0 &i \end{pmatrix} \Cdot \vv{E}_{33} = \begin{pmatrix} -\cos \frac{\pi}{16} \\ -\sin \frac{\pi}{16} \end{pmatrix}
\end{equation}
And finally we project this state back into the \pol{HV} basis in order to evaluate $\gamma$ there:
\begin{align}
   \vv{E}_{\pol{HV}} = \begin{pmatrix} \cos \big( \frac{-3 \pi}{16} \big) &\sin \big( \frac{-3 \pi}{16} \big) \\ -\sin \big( \frac{-3 \pi}{16} \big) &\cos \big( \frac{-3 \pi}{16} \big) \end{pmatrix} \Cdot \vv{E}'_{33} = \frac{-1}{\sqrt{2}} \begin{pmatrix} 1 \\ 1 \end{pmatrix} \, .
\end{align}
Due to the $-1$ factor in front, we see that this wave is 180\degree out of phase with respect to our starting point, in agreement with \figref{fig:sequence3}(b).

Although not shown in Figs~\ref{fig:sequence1}--\ref{fig:sequence3}, we have also verified that the Jones calculus approach properly inverts the sign of the geometric phase when the sense of the path is reversed (that is, when the elements are traversed in reverse order). Just as the order of rotations matters when performing rotations in 3D (the 3D rotation group is non-abelian), the order in which polarization transformations are performed affects the phase shift.

\section{Conclusion}

We have shown that the wave description of geometric phase, based on the composition of waves, provides a more general model than previous theory, which is limited to considering only cyclical paths. The wave model allows us to visualize how $\gamma$ is generated, as the shift in the polarized wave peak location as a wave proceeds through an optical system. Because the wave composition approach is agnostic to paths and treats each polarization state independently, it is possible to analyze the continuous evolution of $\gamma$ as a wave propagates through an optical element, and through any optical system. This approach not only replicates all existing results of cyclical-path geometric phase calculations, it also provides insights that can be used to explain the conditions under which solid-angle-based calculations of $\gamma$ work.

From these new insights, we see that the gauge-dependence of geometric phase plays a critical role in generating phase shifts. As the wave passes from one polarization element to another, if the two elements do not share the same eigenbasis then we see that there can be a sudden phase shift induced by the transition. This explains the need for a ``fourth rule'' for calculating $\gamma$ based on solid angles on the Poincar{\'e} sphere: any consecutive elements sharing an eigenbasis must be treated as a single element for the purposes of tracing geodesics.



Finally, the existing literature has often described geometric phase as exhibiting ``path memory''~\cite{Berry1990a,Anandan1992,Ericsson2003,Carollo2005}. If one considers the path as being the sequence of geodesic curves, then our analysis above has shown that, if we view the path through each successive eigenbasis, then it is not the geodesic curves themselves but rather the basis transformations that occur at path nodes that generate geometric phase. On the other hand, if one considers the physical path of the polarization state, then Figures~\ref{fig:sequence1}\,\&\,\ref{fig:sequence3} demonstrate that this path memory is selective: there may be multiple physical paths between a given cyclical set of nodes which generate the same $\Delta \gamma$ value. Finally, we have shown that the wave model correctly calculates $\gamma$ even though it is entirely agnostic to notions of path. Thus, in this model, geometric phase has no more notion of memory than does propagation phase.


\end{document}